\documentclass[preprint]{aastex}

\begin{document}
\title{Cosmic Shear from Galaxy Spins}
\author{Jounghun Lee}
\affil{Institute of Astronomy and Astrophysics, Academia Sinica,
Taipei, Taiwan}
\email{taiji@asiaa.sinica.edu.tw}
\and 
\author{Ue-Li Pen}
\affil{Canadian Institute for Theoretical Astrophysics,
Toronto, Ont. M5S 3H8, Canada}
\email{pen@cita.utoronto.ca}

\begin{abstract}
We discuss the origin of galactic angular momentum, and the statistics
of the present day spin distribution.  It is expected that the galaxy
spin axes are correlated with the intermediate principal axis of the
gravitational shear tensor. This allows one to reconstruct the shear
field and thereby the full gravitational potential from the observed
galaxy spin fields.  We use the direction of the angular momentum
vector without any information of its magnitude, which requires a
measurement of the position angle and inclination on the sky of each
disk galaxy.  We present the maximum likelihood shear inversion
procedure, which involves a constrained linear minimization.  The
theory is tested against numerical simulations.  We find the
correlation strength of nonlinear structures with the initial shear
field, and show that accurate large scale density reconstructions are
possible at the expected noise level.
\end{abstract}


\newpage 

\newcommand{\beq}{\begin{equation}}
\newcommand{\eeq}{\end{equation}}
\newcommand{\hL}{\hat{L}}
\newcommand{\hT}{\hat{T}}
\newcommand{\bT}{{\bf T}}
\newcommand{\bI}{{\bf I}}
\newcommand{\bL}{{\bf L}}
\newcommand{\bQ}{{\bf Q}}
\newcommand{\hbL}{\hat{\bf L}}
\newcommand{\hbT}{\hat{\bf T}}
\newcommand{\hbQ}{\hat{\bf Q}}
\newcommand{\bC}{{\bf C}}
\newcommand{\hbC}{\hat{\bf C}}
\newcommand{\bx}{{\bf x}}
\newcommand{\bk}{{\bf k}}
\newcommand{\bfv}{{\bf v}}
\newcommand{\br}{{\bf r}}
\newcommand{\by}{{\bf y}}

\section{Introduction}

In the gravitational instability picture of structure formation,
galaxy angular momentum arises from the tidal torquing during the early 
protogalactic stages \cite{hoy49}.  Gravitational torquing results
whenever the inertia tensor of an object is misaligned with the local
gravitational shear tensor.  One might thus expect that the observed 
galaxy spin field would contain some information about the gravitational 
shear field, and possibly allow a statistical reconstruction thereof.  
Although the observational search has yet to detect any significant 
alignments of the galaxy spins so far \cite{han95},  it has been shown 
that the sample size has to be increased in order to detect weak 
alignments (Cabanela \& Dickey 1999 and references therein).  

It was argued that angular momentum only arises at second order 
in perturbation theory \cite{pee69}.  This picture involved a spherical 
protogalaxy, which at first order has no quadrupole moment, and thus cannot 
be tidally torqued.  Later, it was realized that protogalaxies 
should be expected to have order unity variations from spherical symmetry,
and should be tidally torqued at first order \citep{dor70, whi84}. 
However, recent quantitative predictions of this picture 
(Catelan \& Theuns 1996, hereafter CT)
appear to be contradicted by simulations \cite{lem-kau99},  
which leaves the field in a state of confusion.  

The Eulerian angular momentum of a halo in a region $V_E$ relative to its
center of mass in comoving coordinates, 
$\bL^{Eul} = \int_{V_{E}} \rho{\bf x}\times\bfv d^3{\bf x}$, 
can be written in terms of Lagrangian variables to second order accuracy 
using the Zel'dovich approximation:   
\beq
\bL^{\rm 2nd} \propto -\int_{V_L}\bar{\rho}{\bf q}\times
{\bf \nabla}\phi d^3{\bf q},
\label{eqn:l2}
\eeq
where $V_L$ is a Lagrangian counter part of $V_E$, and $\phi$ is 
the gravitational potential.  
Approximating $\nabla\phi$ in equation (\ref{eqn:l2}) by the first three  
terms of the Taylor-series expansion about the center of mass, we obtain 
the first order expression of the angular momentum \cite{whi84} 
in terms of the shear tensor $\bT =(T_{jl})=(\partial_j\partial_l\phi)$ 
and the inertia tensor $\bI =(I_{lk})=(\int q_l q_k d^3{\bf q})$:
\beq
L^{\rm 1st}_{i} \propto \epsilon_{ijk}T_{jl}I_{lk}.
\label{eqn:l1}
\eeq
CT used equation (\ref{eqn:l1}) to calculate the linearly predicted 
angular momentum under the approximation that the principal axes of 
$\bI$ and $\bT$ are uncorrelated.  They also discussed the 
small factor by which the neglect of the correlation 
between $\bI$ and $\bT$ overestimates the angular momentum 
in the context of the Gaussian-peak formalism. In $\S$ 2, we show 
by numerical simulations that this factor in fact is dominant.   

For galaxies, the direction of angular momentum is relatively
straightforward to observe, while the magnitude very difficult.  We
will thus concentrate on the statistics of the direction of the spin,
dealing with $unit$ spin vectors.  We note that
the spin in equation (\ref{eqn:l1}) does not depend on the trace of 
either $\bI$ nor $\bT$.  So we can consider unit trace-free  
tensors, $\hat{T}_{ij}$ ($\hat{T}_{ij}\hat{T}_{ij}=1$), 
and the one point statistics of the spins.
The most general quadratic relation between a unit spin vector 
and a unit traceless shear tensor is 
\beq
\langle \hat{L}_i \hat{L}_j |\bT \rangle \equiv Q_{ij} =
\frac{1+a}{3}\delta_{ij}  - a\hat{T}_{ik} \hat{T}_{kj},
\label{eqn:lij}
\eeq 
where $a \in [0,1]$ is the {\it correlation parameter}, measuring how
well aligned $\bI$ and $\bT$ are.  If $\bI$ and $\bT$ are uncorrelated,  
$a = 1$.  If they are perfectly correlated, $a = 0$.  It is also a
convenient parameterization of the nonlinear effects on the spin-shear 
correlation, and must be measured in numerical simulations.  

\section{Simulations}

We ran the N-body simulations using the PM (Particle-Mesh) code 
described by Klypin \& Holtzman (1997) with $128^3$ particles 
on a $256^3$ mesh in a periodic box of size $L_b = 80 h^{-1} {\rm Mpc}$ for 
the cold dark matter model ($\Omega_0 = 1$, $\sigma_8 = 0.55$ and $h = 0.5$).  
We identified halos in the final output of N-body runs 
(total number of halos, $N_t = 2975$) by the standard 
friends-of-friends algorithm with a linking length of $0.2$.     
The angular momentum of each halo was measured from the positions
and velocities of the component particles 
in the center-of-mass frame.  We also calculated the initial shear 
tensor by taking a second derivative of the gravitational 
potential at the Lagrangian center-of-mass of each  halo.  

Through the simulations, we first tested the validity of the linear 
perturbation theory by calculating the average correlations among the 1st,  
the 2nd order Lagrangian and the Eulerian angular momentum 
obtained from simulations.  
We have found that  
$\langle\hbL^{\rm 2nd}\cdot\hbL^{\rm Eul}\rangle = 0.55$, 
$\langle\hbL^{\rm 1st}\cdot\hbL^{\rm Eul}\rangle = 0.51$, and 
$\langle\hbL^{\rm 1st}\cdot\hbL^{\rm 2nd}\rangle = 0.62$.   
It shows that nonlinear effects only add a factor of $2$ scatter 
to the linearly predicted spin-shear correlation \cite{sug99}.  

Rotating the frame into the shear-principal axes,  we find the optimal
estimation formula for the parameter {\it a} from equation (\ref{eqn:lij}): 
\beq
a = 2-6\sum_{i=1}^{3}\tilde{\lambda_{i}}^{2}|\hL_{i}|^2, 
\label{eqn:a}
\eeq 
where $\{\tilde{\lambda_{i}}\}_{i=1}^{3}$ are the three eigenvalues 
of the trace-free unit shear tensor, satisfying 
$\sum_{i}\tilde{\lambda_{i}}^2 = 1$ and $\sum_{i}\tilde{\lambda_{i}} = 0$. 
Using equation (\ref{eqn:a}) along with the initial shear tensors and the
angular momentum of final dark halos measured from the simulations,  
we calculated the average value of {\it a} to be $0.237\pm 0.023$, 
showing $10\sigma_a$ deviation [$\sigma_a = \sqrt{8 / (5N_t)}$] 
from the value of $0$.       
The inferred value of {\it a} results in a significant correlation of the 
shear intermediate principal axis with the direction of the angular momentum
of halos. We have shown by simulations that the shear and the inertia 
tensors are misaligned with each other to a detectable degree, generating 
net tidal torques even at first order,  
although they are quite strongly correlated 
(it was found by the simulations that 
$\langle I_{11}\rangle=-0.85$, $\langle I_{22}\rangle=0.11$, 
$\langle I_{33}\rangle=0.74$ for the trace free unit inertia tensor
$\bI$ in the shear principal axes frame).   Thus we conclude that the
CT approximations are useful for determining the direction of the spin
vector, but poor for predicting its magnitude \cite{lem-kau99} due to  
the strong correlation between $\bI$ and $\bT$.   The strong
correlation can be understood physically if one considers $\bI$ to be
the collection of particles which have shell crossed, in which case it
is identical to $\bT$.
  
\section{Shear Reconstruction}

As in the CT prescription, the probability 
distribution of $P(\bL|\bT)$ can be described as Gaussian, and   
its directional part, $P(\hbL|\bT)$ is calculated by  
\beq
P(\hbL|\bT) = \int P(\bL|\bT)L^{2}dL = 
\frac{|\hbQ|^{-1/2}}{4\pi}\left( \hat{\bf L}^{\rm
t}\hbQ^{-1} \hbL \right)^{-3/2}
\eeq
where $\bQ$ is given from equation (\ref{eqn:lij}).
According to Bayes' theorem, 
$P(\bT|\hbL)=P(\hbL|\bT) P(\bT)/P(\hbL)$.
Here, $P(\bT)=P[\bT(\bx_1),\bT(\bx_2),\ldots]$ is a joint random 
processes linking different points with each other.  
In the standard picture of galaxy formation from random
Gaussian fields, $P(\bT)$ is given as 
$P(\bT)=N \exp(-\bT^t\bC^{-1}\bT/2)$ where 
$\bC = (C_{ijkl}) = \langle T_{ij}({\bf x})T_{kl}({\bf x}+{\bf r}) \rangle$ 
is the two-point covariance matrix of shear tensors:
\begin{eqnarray}
C_{ijkl}({\bf r}) &=&
(\delta_{ij}\delta_{kl}+\delta_{ik}\delta_{jl}+\delta_{il}\delta_{jk}
)\bigg{\{}\frac{J_3}{6} - \frac{J_5}{10}\bigg{\}} + 
(\hat{r}_i\hat{r}_j\hat{r}_k\hat{r}_l)
\bigg{\{}\xi(r) + \frac{5J_3}{2} - \frac{7J_5}{2}\bigg{\}}
 \nonumber \\ 
&& +  (\delta_{ij}\hat{r}_k\hat{r}_l + \delta_{ik}\hat{r}_j\hat{r}_l
+ \delta_{il}\hat{r}_k\hat{r}_j + \delta_{jk}\hat{r}_i\hat{r}_l
+ \delta_{jl}\hat{r}_i\hat{r}_k + \delta_{kl}\hat{r}_i\hat{r}_j )
\bigg{\{} \frac{J_5}{2}-\frac{J_3}{2} \bigg{\}}.   
\label{eqn:shear}
\end{eqnarray}
Here $\hat{\bf r}={\bf r}/r$, 
$J_n \equiv n r^{-n}\int_0^r \xi(r) r^{n-1} dr$, and 
$\xi(r)=\langle \delta(\bx)\delta(\bx+\br)\rangle$ is the
density correlation function.  

We would like to compute the posterior expectation value, for example
$\langle \bT|\hbL\rangle$.  By symmetry, both the expectation value and
maximum likelihood of $\bT$ occur at $\bT={\bf 0}$, 
which is not the solution we are looking for.  
We must thus consider the constrained expectation,
with the constraint that $\int T_{ij}T_{ij}d^{3}\bx = 1$.  
In the limit that $P(\bT|\hbL)$ is Gaussian, 
this is given by the maximum eigenvector of the posterior
correlation function $\xi_{ijlm}(\bx_\alpha,\bx_\beta)\equiv\langle
T_{ij}(\bx_\alpha) T_{lm}(\bx_\beta)|\hbL \rangle$.  The solution
satisfies 
\beq
\int \xi_{ijlm}(\bx_\alpha,\bx_\beta)
T_{ij}(\bx_\alpha)d^3\bx_\alpha = \Lambda T_{lm}(\bx_\beta)
\label{eqn:tlm}
\eeq
where $\Lambda$ is the largest eigenvalue for which 
equation (\ref{eqn:tlm}) holds.  
In the asymptotic case that $a \ll 1$ as in our simulations results,  
we have a simple expression for the $\hL_i\hL_j$-depending part of the 
posterior correlation function in terms of the traceless shear, 
$\tilde{T}_{ij}=T_{ij}-\delta_{ij}T_{ll}/3$: 
\beq
\tilde{\xi}_{ijlm}(\bx_\alpha,\bx_\beta) = 
- \int\tilde{C}_{ijnk}(\bx_\alpha-\bx_\gamma) 
\tilde{C}_{lmok}(\bx_\beta-\bx_\gamma) 
\hL_n(\bx_\gamma)\hL_o(\bx_\gamma)d^3\bx_\gamma,
\label{eqn:lin}
\eeq
where  $\tilde{\bC}$ is now the trace-free two-point covariance 
matrix of the shear tensors, related to $\bC$ by 
$\tilde{C}_{ijkl} = C_{ijkl} - \delta_{kl}C_{ijnn}/3 
- \delta_{ij}C_{mmkl}/3 + \delta_{ij}\delta_{kl}C_{mmnn}/9$.
Substituting (\ref{eqn:lij}) into (\ref{eqn:lin})
explicitly satisfies a modified version of (\ref{eqn:tlm}).  It is
readily described in Fourier space:
\beq
\int \frac{\tilde{\xi}_{ijlm}(\bk_\alpha,\bk_\beta)}{P(k_\alpha)
P(k_\beta)}\tilde{T}_{ij}(\bk_\alpha)d^3\bk_\alpha = 
\Lambda\tilde{T}_{lm}(\bk_\beta).
\label{eqn:tpk}
\eeq
Equation (\ref{eqn:tpk}) differs from (\ref{eqn:tlm}) by the appropriate 
Wiener filter due to small $a$, since we have assumed no noise.

Since we have discarded the constant
diagonal component which does not depend on $\hL_{i}\hL_{j}$, equation 
(\ref{eqn:lin}) is not positive definite, and in fact has zero trace.  
We can nevertheless use a power iteration to quickly obtain the 
eigenvector corresponding to the largest eigenvalue. One starts with an 
initial guess $\tilde{T}_{ij}^0(\bx_\alpha)$, and defines iterates
$\tilde{T}_{ij}^{n+1/2}=\int\tilde{\xi}_{ijlm}
\tilde{T}^{n}_{lm}$, and $\tilde{T}^{n+1}_{ij} = \tilde{T}^{n+1/2}_{ij}
/\sqrt{\int(\tilde{T}_{ij}^{n+1/2})^2}+\tilde{T}_{ij}^{n-1}$.  
This effectively eliminates the negative eigenvectors from the iteration, 
and converge to the correct solution.  After $n$ iterations, the
fractional error is proportional to $(\Lambda_1/\Lambda_0)^n$, where
$\Lambda_0$ and $\Lambda_1$ are the largest and second largest
eigenvalues, respectively.  In our experience, fifty iterations
converges the result to within a percent of the exact solution.  In
practice, one can expect the correlation to drop off rapidly at large
separations, requiring only an evaluation of a sparse matrix-vector
multiplication.  

The final step is the reconstruction of the density given the
traceless shears.  It is convenient to consider the full shear as a
orthonormal reparametrization in terms of trace and trace-free
components into a vector
$T_i=\{\delta/\sqrt{3},v_2,v_3,\sqrt{2}T_{12},\sqrt{2} T_{23},\sqrt{2}
T_{31}\}$ where $\delta=T_{11}+T_{22}+T_{33}$, 
$v_2=[(-3-\sqrt{3})T_{11}+2\sqrt{3}T_{22}+(3-\sqrt{3})T_{33}]/6$ and
$v_3=[(3-\sqrt{3})T_{11}+2\sqrt{3}T_{22}+(-3-\sqrt{3})T_{33}]/6$.
The two point correlation function (\ref{eqn:shear}) is just linearly
transformed into these new variables, which we will denote $\langle
T_i(\bx) T_j(\bx+\br)\rangle=C^l_{ij}(r)$, where the indices $i,j$ 
run from $1$ to $6$. The inverse correlation will be denoted
${\bf D}=({\bf C}^l)^{-1}$.  From ${\bf D}$ we extract a scalar
correlation $D_{11}(r)$ and a 5 component vector correlation
$E_\mu=\{D_{12},D_{13},D_{14},D_{15},D_{16}\}$.  Similarly, we will
use a Greek index $T_\mu$ to denote the 5 component vector of $T_i$
where $i>1$.  We then obtain the expression for the density
$\langle\delta|T_\mu\rangle=D_{11}^{-1} E_\mu T_\mu$.  If the galaxies
are uniformly spaced on a lattice, one could also use a Fast Fourier
Transform to rapidly perform the shear reconstruction iterations, and the
projection of traceless shear to density. 

\begin{figure}
\plotone{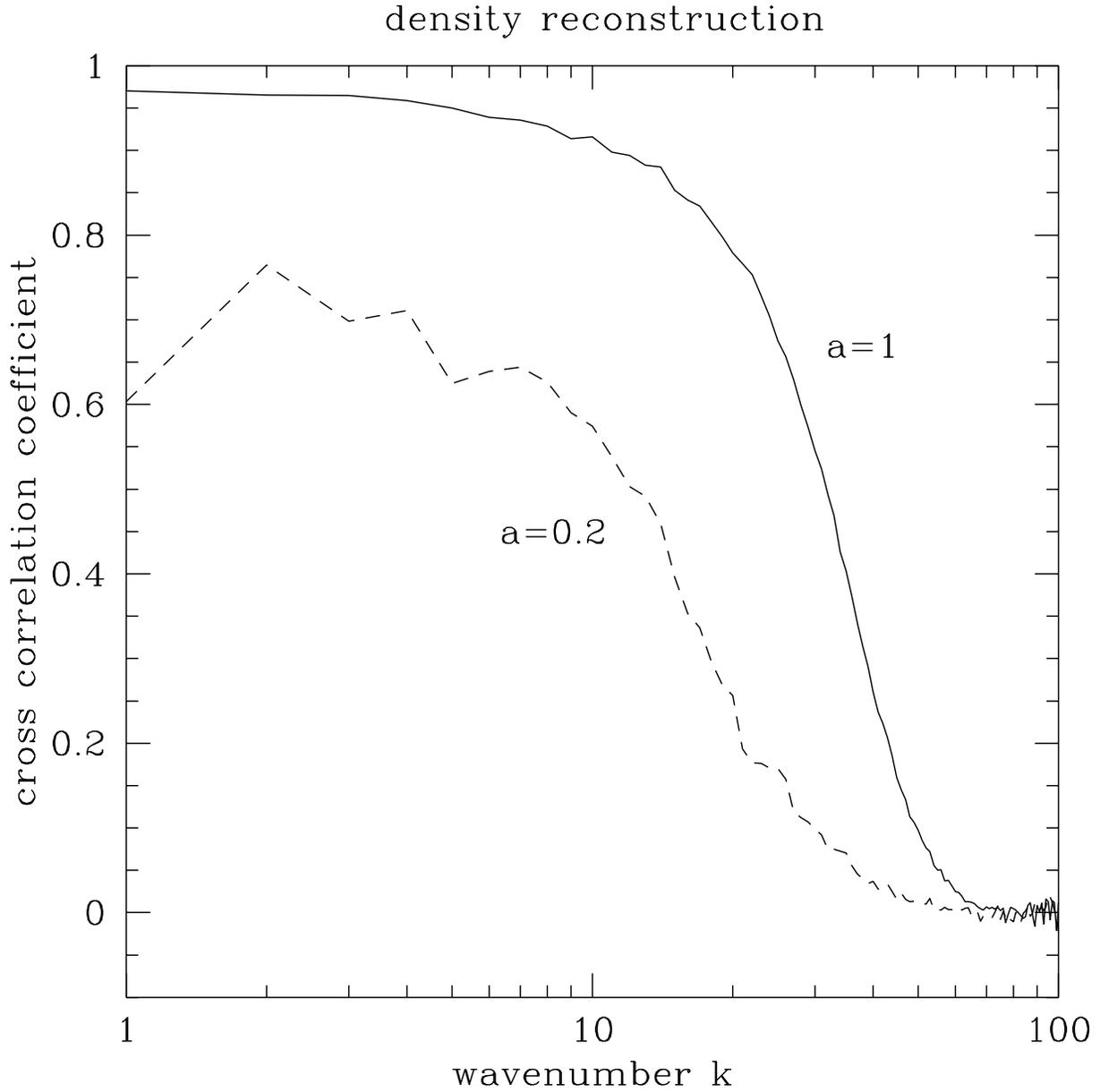}
\caption{Cross correlation coefficient $c=\langle\delta(\bk)
\delta_r(-\bk)\rangle/\sqrt{P(k) 
\langle|\delta_r(\bk)|^2\rangle}$ between original $\delta$ and
reconstructed $\delta_r$ density field in a $128^3$ spin field simulation.  
The correlation parameter $a=1$ refers to the purely random shear model 
\cite{cat-the96}, while $a=0.2$ to the realistic 
shear model as found in our N-body simulations.
The large scale density field is always well reconstructed, while
noise dominates on small scale.\label{fig:ccorr}} 
\end{figure}

To test the reconstruction procedure, we have made a large 
stochastic spin field on a $128^3$ lattice and applied the 
reconstruction procedure with a Wiener filtering scale of $k=32$ 
for the case of a power-law spectrum, $P(k)=k^{-2}$. 
Figure 1 plots the correlation coefficient 
$c=\langle\delta(\bk) \delta_r(-\bk)\rangle/\sqrt{P(k)
\langle|\delta_r(\bk)|^2\rangle}$ between the reconstructed density
field $\delta_r$ and the original density field $\delta$. 
We see that for $a=1$, corresponding to the case of uncorrelated 
shear and inertia tensor, the reconstruction is very accurate 
on large scale (small $k$) and noisy at small scale (large k) as expected.  
Even for the realistic case of $a\sim 0.2$ given the nonlinear effects,  
Figure \ref{fig:ccorr} shows that 
a reconstruction is still quite possible but with less accuracy. 
Note that the reconstructed signal automatically becomes Wiener filtered 
on large scale. 

\section{Observational Effects}

Let us now mention several complications that might arise with real
data, and speculations on how they might be addressed.

The three dimensional spin axis of a disk galaxy can in principle be
determined \cite{han95}, but it requires knowledge of the rotation
curve as well as morphological information, such as extinction maps or
the direction of spiral arms.  A much easier observational task is to
measure only the position angle $\alpha$ on the sky and the projected axis
ratio $r$.  Assuming a flat disk geometry for the light and the $z$
coordinate along the radial direction, this corresponds to observation of 
$r=|{\hL_z}|$ and $\tan(\alpha)=\hL_y/\hL_x$.  We immediately note two
discrete degeneracies: $\hL_z \longrightarrow -\hL_z$ and $\{\hL_x,\hL_y\}
\longrightarrow -\{\hL_x,\hL_y\}$, giving four possible spin
orientations consistent with the observables.  Since we only care
about the spatial orientation of the spin axis, but not its sign, only
a twofold degeneracy exists, which we call $\hbL^a,\hbL^b$.  We can
readily take that into account.  Noting that $P(\hbL^a \vee \hbL^b |T)
= P(\hbL^a|\bT)+P(\hbL^b|\bT)-P(\hbL^a\wedge \hbL^b|\bT)$, the last
term is zero, we simply replace all occurrences of $\hL_i\hL_j$ in
equation (\ref{eqn:lin}) with $\hL^a_i\hL^a_j+\hL^b_i\hL^b_j$. 

The reconstruction itself is quite noisy for an individual galaxy, so 
we will treat as noise the displacement between the observable galaxy 
positions in Eulerian redshift space and their Lagrangian positions 
used in our analysis.  If galaxies are randomly displaced from
their origin by $\sigma_r$, we simply convolve the two point density
correlation function $\xi(r)$ with a Gaussian of variance
$\sigma^2_r$, and appropriately update the shear correlation matrix.
We furthermore note that the use of redshift for distance introduces
an anisotropy in the smoothing Gaussian, which can also be readily
taken into account.  This procedure could in principle be inverted to
measure the nonlinear length scale by measuring $\sigma_r$ if one
knows the intrinsic power spectrum.  One could vary $\sigma_r$ until
one maximized the eigenvalue $\Lambda$ in Equation (\ref{eqn:tlm}).

The shear reconstruction correlator (\ref{eqn:lin}) is only accurately
defined at each galaxy position, so the integral is should be replaced
by a sum over galaxies.  The eigenvector iteration scheme works
the same way with the discretized matrix.  The density is recovered
from the trace-free shear in the same fashion as the continuum limit,
where again we only reconstruct the density at the galaxy positions.
The density at any other point is reconstructed in a similar fashion
using the posterior expectation value given the Gaussian random field.

The observed spin vectors are of the luminous materials.  They can be
observed out to large radii, tens of kpc, through radio emission of
the gas.  While some warping of the disks are seen, the direction of 
the spin vector generally changes only very modestly as one
moves to larger radii.  This suggests that the luminous angular
momentum be well correlated with the angular momentum of the
whole halo.  The standard galaxy biasing is not expected to affect
the reconstruction, since we have not used the galaxy densities but
only the spin directions.  Merging of galaxies is also taken into 
account by this reconstruction procedure, since the resulting spin 
vectors are expected to align with the constituent orbital angular 
momentum vectors. The latter is again predicted by the same shear 
formula.  The merger effects are included in the N-body simulations
presented above.

\section{Conclusions}

We have presented a direct inversion procedure to reconstruct the
gravitational shear and density fields from the observable unit galaxy
spin field.  The procedure is algebraically unique, dealing only
with linear algebra, and is computationally tractable when done
iteratively.  We have shown that the angular momentum-shear
correlation signal in simulations is strong enough to allow useful
reconstructions.  Direct N-body simulations suggest that nonlinear
collapse effects only reduce the linearly predicted spin-shear
correlation by a factor of 2.  The deprojection degeneracy can be
incorporated in a straightforward fashion.  Large galaxy surveys with
a million galaxy redshifts are coming on line soon.  This procedure
may allow a density reconstruction which is completely independent of
any other method that has thus far been proposed.  We have
demonstrated its effectiveness in simulated spin catalogs.  Due to the
nature of this letter, only the key results essential to the 
reconstruction algorithm have been presented. 
Detailed intermediate steps and justifications will be shown
elsewhere \cite{lp00}.

\acknowledgments 
We thank Giuseppe Tormen and Antonaldo Diaferio for providing the 
halo-finding algorithm, and the referee for useful suggestions. 
J. Lee also thanks Anatoly Klypin for helpful 
comments on the use of the PM code. 
This work has been supported by Academia Sinica, and partially by 
NSERC grant 72013704 and computational
resources of the National Center for Supercomputing Applications.

\end{document}